# Screening potential topological insulators in half-Heusler compounds via compressed-sensing

Jianghui Liu, Guohua Cao, Zizhen Zhou, Huijun Liu[*]

*Key Laboratory of Artificial Micro- and Nano-Structures of Ministry of Education and School of Physics and Technology, Wuhan University, Wuhan 430072, China*

Ternary half-Heusler compounds with widely tunable electronic structures, present a new platform to discover topological insulators. Due to time-consuming computations and synthesis procedures, the identification of new topological insulators is however a rough task. Here, we adopt a compressed-sensing approach to rapidly screen potential topological insulators in half-Heusler family, which is realized via a two-dimensional descriptor that only depends on the fundamental properties of the constituent atoms. Beyond the finite training data, the proposed descriptor is employed to screen many new half-Heusler compounds, including those with integer and fractional stoichiometry, and a larger number of possible topological insulators are predicted.

## I. INTRODUCTION

Topological insulators (TIs), which exhibit metallic surface states inside the bulk band gap, have attracted considerable research interests in condensed-matter physics for their potential applications in spintronics and quantum computing [1-10]. This novel quantum form of materials is first predicted and realized in binary compounds derived from the semimetal HgTe [11, 12]. Recently, various studies of TIs have been extended to ternary systems [13-20]. It is known that half-Heusler compounds are ternary solids which in general consist of 8 or 18 valence electrons. Based on the band structure calculations, many of them are proved to be topologically nontrivial, which have similar electronic properties to the binary HgTe [21].

For a given compound system, it is usually complex to identify its topological nature since detailed calculations of the electronic structure and $Z_2$ invariant are required. Hence, a brief rule of thumb is desperately desired to distinguish topological systems from trivial ones. Recently, several high-throughput empirical descriptors or symmetry-based indicators are developed for predicting various TIs [22-26]. However, they unfavorably necessitate detailed band structure calculations or are limited by systems already known in databases. To accelerate the process of discovering novel materials, many advanced approaches are utilized, including data mining, artificial neural networks, and etc. [27-31]. Nevertheless, the interplay between a large amount of training data and overwhelming combinations of functional model is still abstruse. It is thus necessary to find out a physically meaningful descriptor such that the relation between the data and the trained models can be easily interpreted. In this respect, the compressed-sensing-based approach named SISSO (Sure Independence Screening and Sparsitying Operator) has been proved to be a promising and efficient tool to derive descriptors from a small set of training data, as demonstrated for many material properties such as stability of perovskite oxides and topological transitions in two-dimensional (2D) materials [32-35]. Specifically, given a vector $P$ consisting set of materials of various properties and a sensing matrix with element $D_{i,j}$ formed by the linear projection of $i$-material into the $j$-feature, the descriptor can be extracted from the sparse vector $c$ by solving $Dc = P$ [36].

---

[*] Author to whom correspondence should be addressed. Electronic mail: phlhj@whu.edu.cn

In this work, a 2D descriptor based on compressed-sensing method is proposed to rapidly screen potential TIs in half-Heusler compounds. The descriptor only depends on the atomic number, the Pauling electronegativity, and the valence electron number of the constituent atoms, rather than the material itself. Beyond a small collection of training data, the optimized descriptor is further generalized to identify the topological nature of a large number of new candidate compounds in the half-Heusler family, including those with fractional stoichiometry.

## II. METHODOLOGY

We employ the SISSO approach developed by Ouyang *et al* [36] to acquire an efficient descriptor for predicting topological nature. The method relies on two key steps: the first one is feature-space construction by building analytical functions of the input variables, which are usually the fundamental properties of the constituent atoms such as the atomic number $Z$, the principal quantum number $n$, the affinity energy $EA$, the relative atomic mass $M$, the Pauli electronegativity $\chi$, and the number of valence electrons $VE$. The combinations of these primary features are then established by iteratively applying the chosen algebraic operators $\{I,+,-,\times,/,\exp,\log,|-|,\sqrt{},^{-1},^{2},^{3}\}$ up to a certain complexity cutoff. The second step is the descriptor identification by SISSO, where the SIS selects features $D_n$ based on combinations of target properties and the SO seeks the optimal $\Omega$-dimension features that minimizes the overlap among convex hulls enveloping subsets of data [37]. Progressively, larger dimensionality $\Omega$ is tested until yielding perfect classification of all data in the training set. In the present work, $\Omega = 2$ is found to be sufficient to classify the training data of 116 half-Heusler compounds, whose topological natures have been theoretically calculated and/or experimentally confirmed. The so-called leave-one-out cross-validation (LOOCV) is then used to quantify the predictive ability of the proposed descriptor. To confirm the reliability of the SISSO identified descriptor, first-principles calculations are performed for several example systems, which are based on the projector augmented-wave (PAW) method [38, 39] within the framework of density functional theory (DFT) [40-42]. The Perdew-Burke-Ernzerhof (PBE) with the generalized gradient approximation (GGA) is used for the exchange and correlation functions [43]. The plane-wave cutoff of 500 eV and a 13×13×13 Monkhorst-pack ***k*** mesh [44] are adopted for the Brillouin zone integrations. In addition, the spin-orbit coupling (SOC) is explicitly considered in the calculations.

## III. RESULTS AND DISCUSSIONS

As known, the electronic and topological properties of a given system is closely related to its crystal structure. Figure 1 shows a ball-and-stick model of the half-Heusler compound, which possesses a face-centered-cubic (fcc) structure described by the space group F-43m. The compound can be identified by the chemical formula ABC, where A, B and C atoms occupy Wyckoff 4*b* (1/2, 1/2, 1/2), 4*c* (1/4, 1/4 ,1/4), and 4*a* (0, 0, 0) atomic positions, respectively [45]. Normally, the 4a and 4b atoms form the ionic NaCl-type substructure, while the 4a and 4c construct the covalent ZnS-type one. Within such structure type, 116 half-Heusler compounds are taken into account which include 58 topologically nontrivial systems and 58 trivial insulators [14, 46-54]. It should be noted that the bulk insulating gap of the topologically ordered half-Heusler systems can be created by applying appropriate strain [46], and the topological natures of these 116 compounds have been confirmed by experimental measurement and/or theoretical calculations.

Using the compressed-sensing approach, we pinpoint the optimal one from 20,000 potential descriptors, which is constructed by appropriately combining the fundamental properties of the constituent atoms A, B and C:

$$D_1 = \frac{(Z_B^2 + Z_A \times Z_C) \times e^{\chi_C}}{VE_A \times VE_B}, \tag{1}$$

$$D_2 = \ln(Z_B) \times Z_B \times VE_A \times |Z_B - |Z_A - Z_C||. \tag{2}$$

Note here only three of the primary features are included, which are the atomic number $Z$, the Pauli electronegativity $\chi$, and the number of valence electrons $VE$. Using $D_1$ and $D_2$ as coordinates, Figure 2 plots the "phase diagram" of the above-mentioned 116 half-Heusler compounds, where we see that the topologically nontrivial and trivial systems can be perfectly classified by the two-dimensional descriptor. The boundary between the two domains can be determined by a dividing line with $D_2 = 156.78 D_1 - 278383.85$, which is calculated using the support-vector machine (SVM) technique [55]. It is known that the topology of band structure is closely related to the relative positions of the $\Gamma_6$ and $\Gamma_8$ states. The energy difference between them ($\Delta = \Gamma_6 - \Gamma_8$) would be negative for topologically nontrivial systems and positive for trivial cases. Unlike the (Bi, Sb)$_2$(Se, Te)$_3$ systems, the nontrivial half-Heusler compounds exhibit an *s-p* band inversion no matter SOC is considered or not, which is attributed to the mass-velocity and Darwin corrections [48]. According to the expression of the $D_1$ descriptor, a system tends to exhibit TI nature if it has larger atomic number ($Z_A$, $Z_B$, $Z_C$). This is reasonable since the energy of the *s* valence electrons could be weakened in heavy atoms due to relativistic effects [48], which in turn leads to more negative $\Delta$ value. In addition, we see that larger $\chi_C$ could also give a larger value of $D_1$, since the anions with strong electronegativity tends to narrow the gap between $\Gamma_6$ and $\Gamma_8$ states ($\Delta$) and thus it is easier to exhibit topologically nontrivial behavior [33, 53]. It should be noted that the variables $\chi_C$ and $Z_C$ are both included in the $D_1$ so that they cannot be simultaneously maximized. If the atom C is selected from group VIA or VIIA, one may obtain relatively bigger values of $\chi_C$ and $Z_C$, and thus a much larger value of $D_1$ is expected. As a consequence, the system could have larger chance to exhibit TI nature, which is consistent with previous DFT calculations [53]. Besides, it was suggested that the electronic properties of half-Heusler compounds are sensitive to the valence electron count [56]. To go further, Eq. (1) indicates that a half-Heusler compound would exhibit TI nature if it has smaller $VE_A$ and/or smaller $VE_B$. On the other hand, if we focus on the region with $1886 < D_1 < 2110$ (enclosed by two vertical lines of the inset), we find a system with smaller value of $D_2$ tends to fall into the topologically nontrivial domain. This is reasonable since here $|Z_A - Z_C|$ are always smaller than or equal to $Z_B$ and Eq. (2) can be thus simplified to:

$$D_2 = \ln(Z_B) \times Z_B \times VE_A \times (Z_B - |Z_A - Z_C|). \tag{3}$$

Among the 7 half-Heusler compounds within this region, 4 of them (ScPtBi, LuPdBi, HoPdBi, KAlGe) exhibit smaller $Z_B$ and/or bigger $|Z_A - Z_C|$ values and thus have smaller $D_2$. In contrast, 3 trivial systems (YPtSb, LuAuSn, ZrIrBi) possess larger $Z_B$ and/or smaller $|Z_A - Z_C|$ values.

To check the robustness of our proposed descriptor, we have performed the so-called

LOOCV. That is, one sample is randomly removed from the 116 half-Heusler compounds and the remaining systems are used as training data to generate a new optimized descriptor. In 98.3% of the times, we find that the LOOCV reproduces the same descriptor obtained from the whole data. Such a fact suggests that the SISSO identified descriptor is rather robust, which essentially stems from the profound physical meaning discussed above.

Beyond the 116 training set, the proposed 2D descriptor can be utilized to theoretically predict other potential TIs in the half-Heusler family. As demonstrated before [57], 481 new half-Heusler compounds out of 71178 possible compositions are predicted to be likely stable, among which many topologically nontrivial materials could be found. The result is illustrated in Figure 3, where 100 new compounds appeared on the right side of the dividing line are quickly predicted to be topologically non-trivial. To check the predictive power, we have done additional first-principles calculations on 6 randomly selected compounds in Fig. 3, which are marked by colored triangles. Among them, 3 systems (LiTeAu, MgAuPb, and LiPbIn) come from topologically non-trivial region and the others (TiNiPb, TiIrSb, and TaRuBi) from trivial domain, which are consistent with our first-principles calculations. For example, the calculated band structure of LiTeAu shown in Figure 4(a) exhibits obvious band inversion between the $\Gamma_6$ (dominated by *s*-orbital) and $\Gamma_8$ sates (governed by *p*-orbital), which characterizes its topologically non-trivial feature. In contrast, the band structure of TiIrSb shown in Fig. 4(b) exhibits a normal band order with a gap of 0.97 eV. These two prototypical examples thus provide direct evidence of strong predictive power of our 2D descriptor. The screened 100 compounds in the non-trivial region suggest that there is still room to discovery new ternary TI candidates in the half-Heusler family, which need further theoretical and experimental investigations.

In addition to the above-mentioned ternary systems with integer stoichiometry, many half-Heusler compounds may contains four or more elements and thus have fractional stoichiometry. It is interesting to check the topological nature of these systems, which is rather difficult from first-principles calculations since very large supercell is needed.

Such a fundamental question can be readily addressed by utilizing our 2D descriptor. For example, if we consider a particular half-Heusler compound with nominal formula $Sc_xY_yLu_{1-x-y}Cu_aAg_bAu_{1-a-b}Ge_mSn_nPb_{1-m-n}$ (Here the stoichiometry $x, y, a, b, m, n$ varies in the ranges of 0~1). The $(D_1, D_2)$ values of such a complex system can still be calculated from Eq. (1) and (2) if we define three weighted atomic numbers,

$$Z_A = xZ_{Sc} + yZ_Y + (1-x-y)Z_{Lu}, \qquad (4)$$

$$Z_B = aZ_{Cu} + bZ_{Ag} + (1-a-b)Z_{Au}, \qquad (5)$$

$$Z_C = mZ_{Ge} + nZ_{Sn} + (1-m-n)Z_{Pb}, \qquad (6)$$

three weighted Pauli electronegativity,

$$\chi_A = x\chi_{Sc} + y\chi_Y + (1-x-y)\chi_{Lu}, \qquad (7)$$

$$\chi_B = a\chi_{Cu} + b\chi_{Ag} + (1-a-b)\chi_{Au}, \qquad (8)$$

$$\chi_C = m\chi_{Ge} + n\chi_{Sn} + (1-m-n)\chi_{Pb}, \qquad (9)$$

and three weighted valence electrons,

$$VE_A = xVE_{Sc} + yVE_Y + (1-x-y)VE_{Lu}, \qquad (10)$$

$$VE_B = aVE_{Cu} + bVE_{Ag} + (1-a-b)VE_{Au}, \qquad (11)$$

$$VE_C = mVE_{Ge} + nVE_{Sn} + (1-m-n)VE_{Pb}. \qquad (12)$$

If each stoichiometry varies at an interval of 0.1, one can artificially produce $\sim 1.95 \times 10^5$ different half-Heusler compounds, including the YAuPb with integer stoichiometry. Similar procedures can be done for other three typical half-Heusler compounds (BaPtS, HfCoSb, and KAlGe) with fractional stoichiometry. In total, $\sim 7.8 \times 10^5$ $(D_1, D_2)$ pairs for the four kinds of mother system are obtained and then mapped into the original phase diagram, as indicated in Figure 5. According to the above discussions, those systems falling into the blue domain ($\sim 3.89 \times 10^5$) could be potential TIs, which await future theoretical or experimental confirmations.

**IV. SUMMARY**

In summary, we have applied the compressed-sensing method to derive a 2D high-throughput descriptor for screening possible TIs in the half-Heusler compounds. The descriptor is merely defined by the atomic number, the valence electron number, and the Pauli electronegativity of the constituent atoms. The reliability and predictive power of the proposed descriptor is demonstrated by explicit first-principles calculations. Utilizing such an efficient descriptor, we have discovered 100 topologically nontrivial compounds in the half-Heusler family, most of which are less studied before. Furthermore, the topological nature of half-Heusler systems with fractional stoichiometry could also be readily identified, which is of particular interest for the related experiments. Although with very simple 2D form, our proposed descriptor has profound physical insight and is expected to be generalized to other similar systems.

**ACKNOWLEDGMENTS**

We thank financial support from the National Natural Science Foundation of China (Grant Nos. 51772220 and 11574236). The numerical calculations in this work have been done on the platform in the Supercomputing Center of Wuhan University.

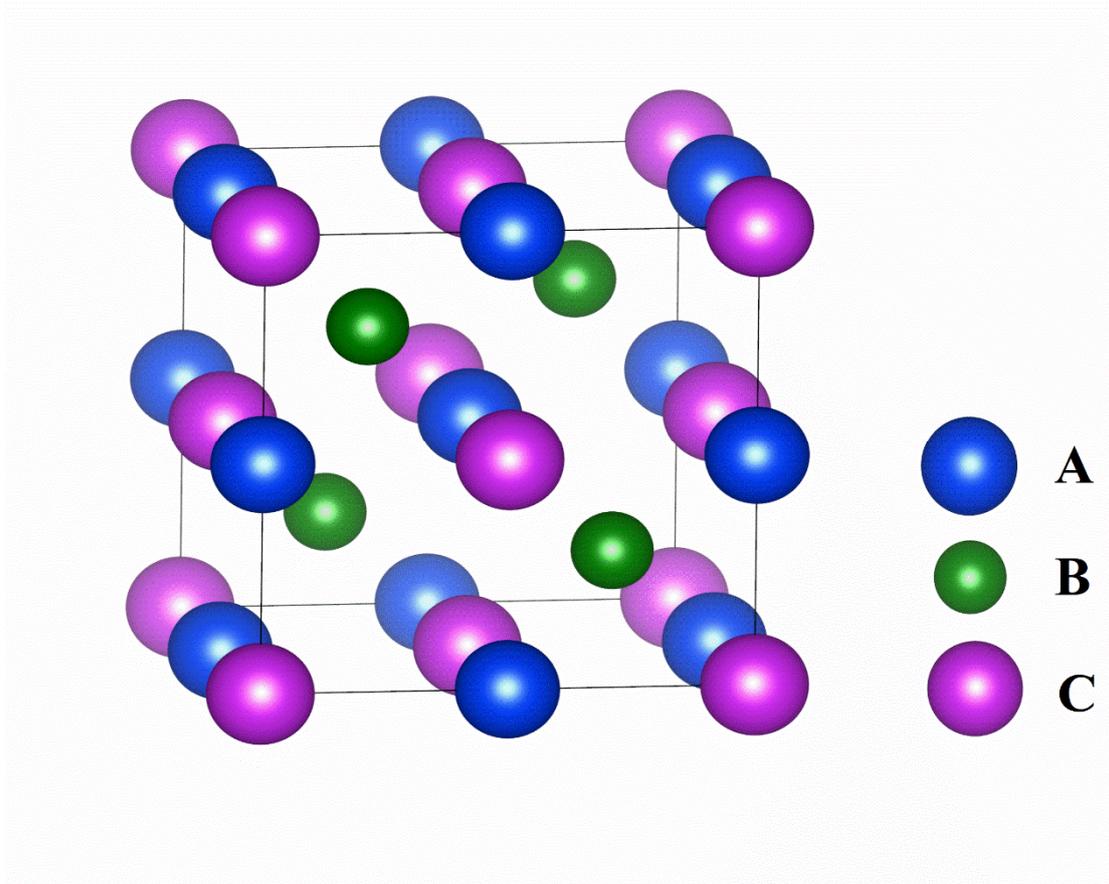

**Figure 1.** The crystal structure of half-Heusler compounds with nominal formula ABC.

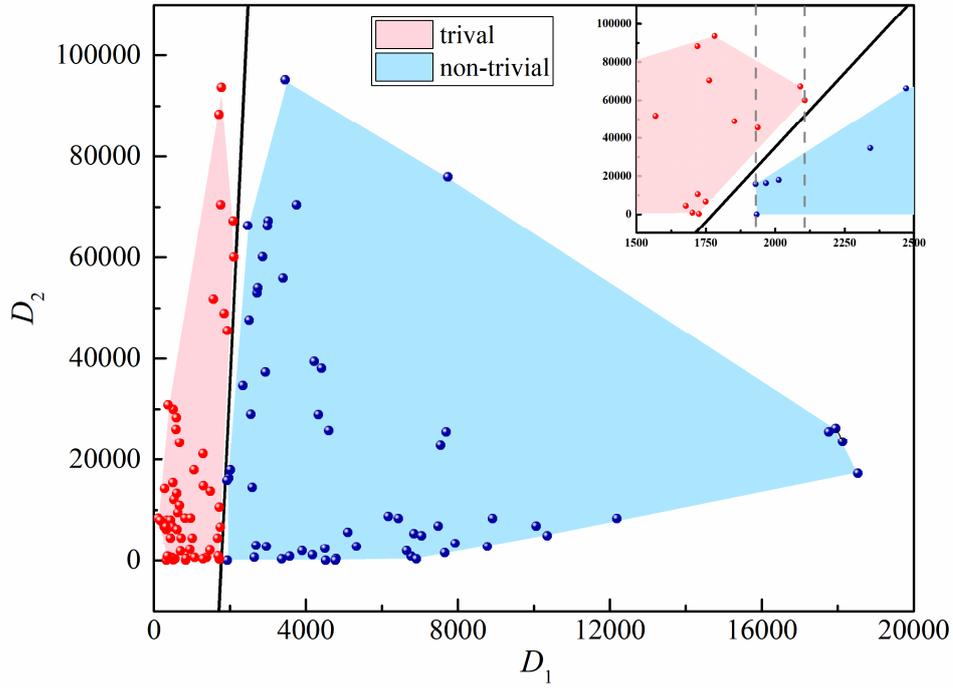

**Figure 2.** Phase diagram of the 116 training data defined by the 2D descriptor. Topologically trivial and non-trivial systems are respectively denoted as the red and blue areas, which are determined by connecting the outermost red and blue points. The inset shows the boundary area between topologically trivial and non-trivial region. The two vertical dashed-lines correspond to $D_1 = 1886$ and $D_1 = 2110$, respectively.

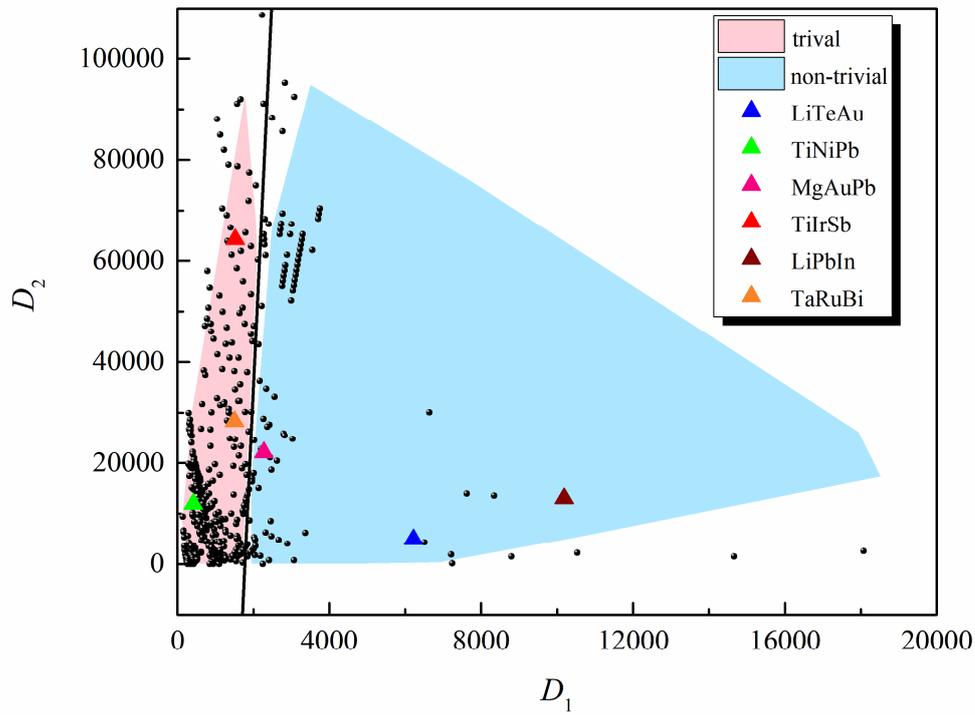

**Figure 3.** The mapping of 481 new half-Heusler compounds (black dots), as determined by the 2D descriptor. The colored triangles indicate 6 example systems whose topological natures are further checked by first-principles calculations.

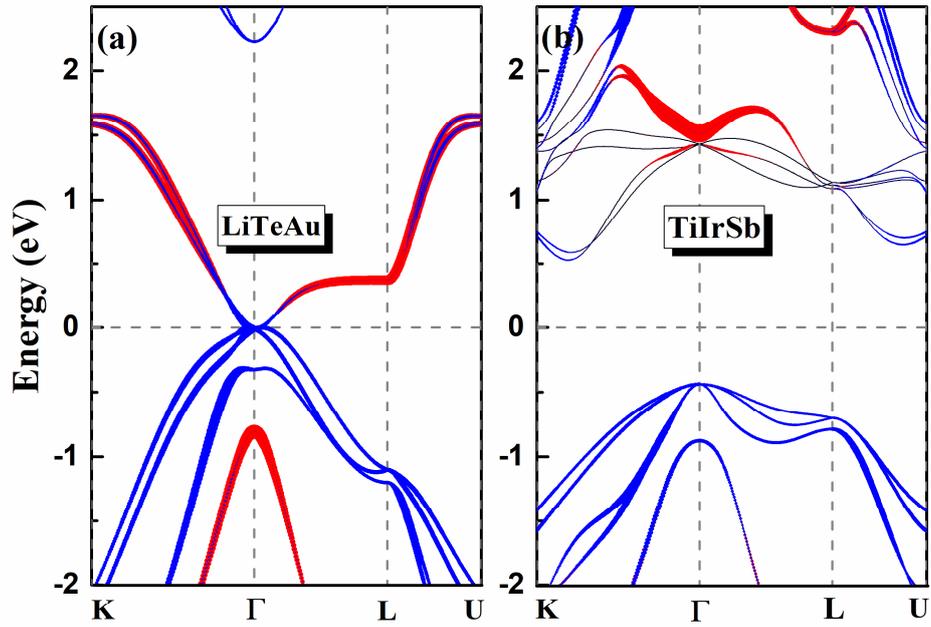

**Figure 4.** Band structures of (a) LiTeAu and (b) TiIrSb. The Fermi level is at 0 eV. The red lines represent the *s*-orbital composition, while blue lines correspond the *p*-orbital composition, and the sizes are proportional to their contributions.

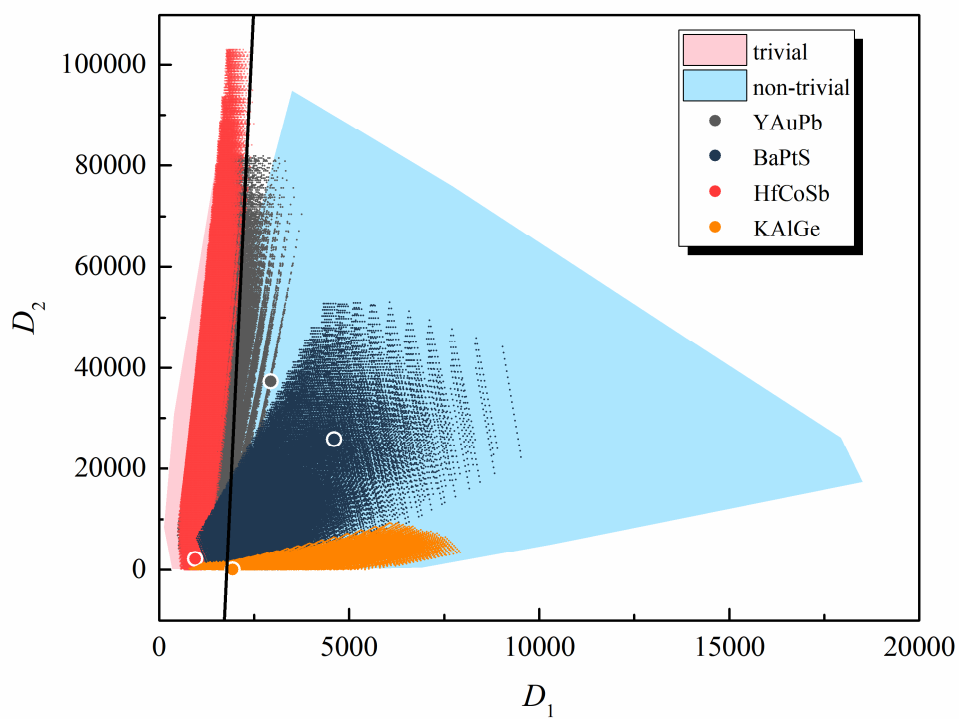

**Figure 5.** The mapping of ~7.8 ×10$^5$ half-Heusler compounds with arbitrary mutation of atoms and/or stoichiometry, as determined by the 2D descriptor. The colored points indicate 4 compounds with integer stoichiometry (YAuPb, BaPtS, HfCoSb and KAlGe).